\documentclass[aps,pra,amsmath,amssymb,superscriptaddress,reprint]{revtex4-1}

\usepackage{graphicx}
\usepackage{dcolumn}
\usepackage{bm}
\usepackage[utf8]{inputenc}
\usepackage[T1]{fontenc}
\usepackage{booktabs, array, mathptmx, mathrsfs, float, tabularx, booktabs, lipsum, amsmath, multirow}
\usepackage{siunitx, xcolor}
\usepackage{pifont}
\usepackage[version=4]{mhchem}
\usepackage[colorlinks,linkcolor=blue,anchorcolor=blue,citecolor=blue]{hyperref}
\usepackage[linesnumbered,ruled,vlined]{algorithm2e}

\begin{document}


\title{Variational quantum circuit learning of entanglement purification in multiple degrees of freedom}

\author{Hao Zhang}
\affiliation{Purple Mountain Laboratories, Nanjing 211111, China}

\author{Xusheng Xu}
\email{thuxuxs@163.com}
\affiliation{Department of Physics, State Key Laboratory of Low-Dimensional Quantum Physics, Tsinghua University, Beijing 100084, China}

\author{Chen Zhang}
\affiliation{Purple Mountain Laboratories, Nanjing 211111, China}

\author{Man-Hong Yung}
\affiliation{Shenzhen Institute for Quantum Science and Engineering, Southern University of Science and Technology, Shenzhen 518055, China}

\author{Tao Huang}
\email{htao@bupt.edu.cn}
\affiliation{Purple Mountain Laboratories, Nanjing 211111, China}
\affiliation{State Key Laboratory of Networking and Switching Technology, Beijing University of Posts and Telecommunications, Beijing 100876, China}

\author{Yunjie Liu}
\affiliation{Purple Mountain Laboratories, Nanjing 211111, China}
\affiliation{State Key Laboratory of Networking and Switching Technology, Beijing University of Posts and Telecommunications, Beijing 100876, China}

\date{\today}

\begin{abstract}
Entanglement purification is a crucial technique for promising the effective entanglement channel in noisy large-scale quantum networks, yet complicated in designing protocols in multi-degree of freedom (DoF). To execute the above tasks easily and effectively, developing a learning framework for designing the entanglement purification with multi-DoF is a promising way and still an open research question. Inspired by variational quantum circuit (VQC) with remarkable advantage in learning optimal quantum operations with near-term quantum devices, in this paper we propose an effective VQC framework for the entanglement purification in multi-DoF and exploit it to learn the optimal purification protocols of the objective function which are based on postselection. By properly introducing additional circuit lines for representing each of the ancillary DoFs of all the particles, e.g., space and time, the parametrized quantum circuit can effectively simulate scalable entanglement purification. To verify our framework, the well-known protocols in linear optics are learned well with alternative operations in low-depth quantum circuit. Moreover, we simulate the multipair cases with multi-DoF to show the scalability and discover one-round protocols. Our work provides an effective way for exploring the entanglement purification protocols in multi-DoF and multipair with near-term quantum devices.
\end{abstract}


\maketitle


\section{Introduction}
Quantum entanglement which shows the nonlocal correlation between two or more objects is an intriguing phenomenon in quantum mechanics and has no classical counterpart \cite{entanglement}. One usually uses quantum entanglement as a crucial resource for building quantum channel in quantum networks \cite{quantumnetwork,QInternet1,QInternet2,QInternet3,QInternet4}. However, in practice, entanglement is so fragile in a noisy environment that it is hard to be used directly as an effective quantum channel. The reason is that under the influence of noise, a pure maximally entangled state becomes a mixed one. To overcome this problem, a technique called entanglement purification is proposed to improve the fidelity of the damaged entangled state \cite{BBPSSW,DEJMPS,Bmixed,pannature1,simonpan,lionestep,shengonestep,shengdeng,rendeng,wangdeng,zhangdeng,jiangEPP,
pannature2,Wineland,pannp,Hanson,husheng,ursinprl,shengsb,ursinprapplied,microwaveepp,scpma}.
The first entanglement purification protocol utilizes another copy of mixed entangled state in Werner form as an auxiliary ``target'' state and executes bilateral controlled-NOT (CNOT) operations and parity check to acquire the information of ``source'' pair \cite{BBPSSW}. Subsequently, the protocol is developed without requirement of Werner form and has higher efficiency in recursive procedure \cite{DEJMPS}. The above protocols are based on CNOT gates between two entangled pairs. It is hard to accomplish this operation in experiment especially for photons. Therefore, in optical systems, the feasible way is to bring in ancillary photonic degree of freedom (DoF), such as space and time. The first photonic entanglement purification protocol which makes use of ancillary DoF is based on selecting the spatial modes of entangled pairs \cite{pannature1} and subsequently performed in experiment  \cite{pannature2}. Besides, one can design entanglement purification protocols with only one pair of photons with multi-DoF hyperentanglement \cite{simonpan,lionestep,shengonestep,shengdeng,rendeng,wangdeng,husheng,ursinprl,shengsb,ursinprapplied}, e.g. Simon-Pan \cite{simonpan} and  Hu-Huang-Sheng-Zhou \emph{et al.} (HHSZ+) \cite{husheng} protocols. Up to now, many interesting purification protocols have been proposed for various cases but lack of a learning framework. As the number of entangled pair and DoF increases, the design of entanglement purification becomes more complicated and challenging.

In recent years, machine learning has been considered for processing quantum information \cite{QML,MLquantum,MLphy,NISQA}. Some basic protocols in quantum communication, such as quantum teleportation \cite{teleportation}, entanglement purification, and quantum repeaters \cite{QRBDCZ}, are discovered by classical agents \cite{MLEPP}. Quantum gate operations play a key role in quantum information processing. Therefore, compared with classical machine learning, an approach directly optimizing quantum gates called variational quantum circuit (VQC) has its inherent advantage for handling quantum information tasks \cite{VQC1,VQC2,VQC3}. The VQC learning has been widely applied in various areas, such as quantum computing and quantum chemistry, for its remarkable advantage in training with near-term quantum devices. This inspires us that VQC learning may be an effective way to design the protocols of entanglement purification in multi-DoF. Recently, entanglement purification has been performed as a simple instance in local operation and classical communication framework based on parameterized quantum circuits \cite{VQCEPP}. However, directly simulating operations on two pairs is limited in some cases, such as the difficult CNOT operation of photons. Therefore, developing an effective VQC learning framework for entanglement purification including multi-DoF has practical significance but there is still no research.

In this paper, we propose an effective VQC learning framework for the entanglement purification in multi-DoF and exploit it to learn the optimal purification protocols of the objective function which is based on postselection. Additional quantum circuit lines are introduced in VQC to represent the high-dimensional DoFs of particle. In our VQC learning framework, the parameterized ansatz part plays the central role in learning the local quantum operations of entanglement purification. The classical communications used for exchanging the information between two users are included in measurement part. As examples, the well-known linear optical entanglement purification protocols, including Pan-Simon-Brukner-Zeilinger (PSBZ) \cite{pannature1}, HHSZ+ \cite{husheng}, Simon-Pan \cite{simonpan} and etc, are learned well and the different operations of entanglement purification are discovered automatically. To verify the scalability of our framework, the cases of multi-pair with multi-DoF are also performed well and the results indicate that multi-pair entangled states can be purified by a one round way that is different from the conventional recursive process. Moreover, entanglement purification is also simulated with different noisy channels. Our framework provides an alternative way to understand and design the entanglement purification in multi-DoF by variational quantum learning and has extensive applications for the other areas of quantum information, such as quantum networks.

The article is organized as follows: In Sec.\ref{secVQC}, we introduce our VQC learning framework for the entanglement purification in multi-DoF. In Sec. \ref{secLEP}, the well-known protocols of the entanglement purification with multi-DoF are learned. In Sec. \ref{secscalability}, the cases of multi-pair with multi-DoF are verified. In Sec. \ref{secnoise}, the noisy channels are considered in simulations.  Section \ref{secDS} covers discussion and summary.

\begin{figure}[]
\includegraphics[width=\linewidth]{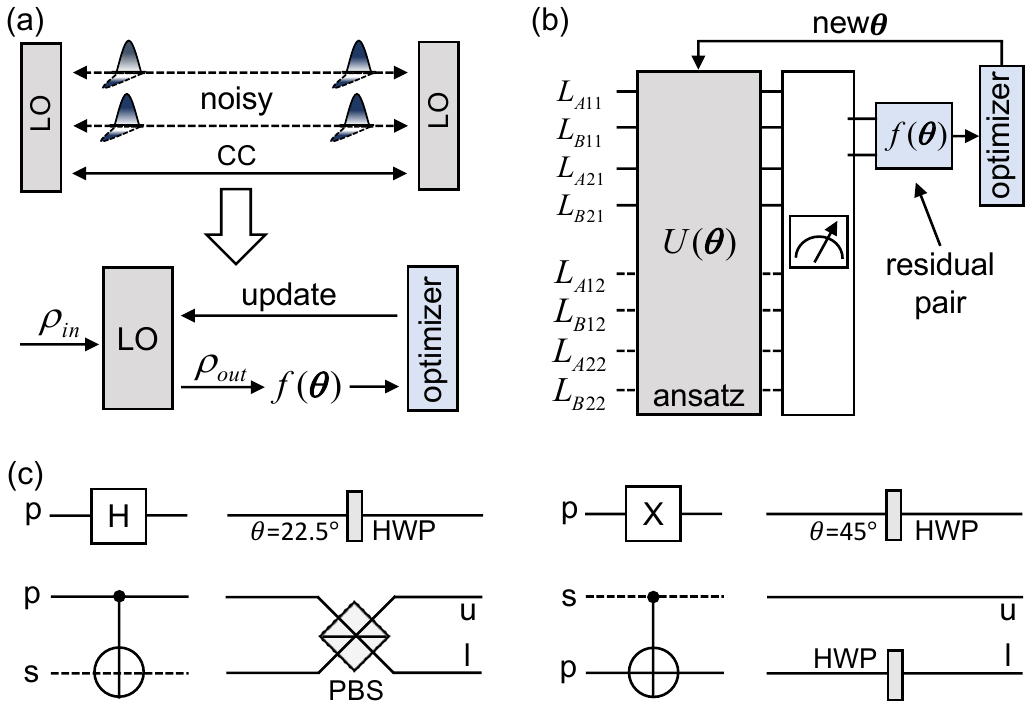}
\caption{The VQC learning framework of the entanglement purification with multi-DoF. (a) The simple schematic diagrams of the entanglement purification in optical systems and VQC. LO, local operations; CC, classical communications. (b) An instance for the VQC of the entanglement purification with polarization and spatial DoFs. (c) The physical implementations of some basic quantum gates in linear optics. The meanings of symbols are: p (s), polarization (spatial) DoF; u (l), upper (lower) spatial DoF; H, Hadamard gate; X, Pauli-X gate; HWP, half-wave plate; PBS, polarizing beam splitter.}\label{VQCEPP}
\end{figure}

\section{VQC framework for the entanglement purification in multi-DoF}\label{secVQC}
A pure entangled state labeled with $\rho_{\text{pure}}$ will become a mixed one $\rho_{\text{mixed}}$ when it is distributed over a noisy quantum channel. However, entanglement purification is an inverse process which can improve the fidelity of a mixed state. Those two processes can be described by the formula
\begin{eqnarray}\label{noiseEPP}
\rho_{\text{pure}}\stackrel{\text{noise}}{\longrightarrow}\rho_{\text{mixed}}\stackrel{\text{EP}}{\longrightarrow}\rho_{\text{purified}}.
\end{eqnarray}
Here, EP is entanglement purification, and $\rho_{\text{purified}}$ is the density matrix of a purified mixed state and can equal to $\rho_{\text{pure}}$ after a perfect entanglement purification protocol. In long-distance quantum channel, an entangled state is usually shared by two nonlocal quantum nodes shown in the upper part of Fig. \ref{VQCEPP} (a). Because of non-locality, the two users are only allowed to execute entanglement purification using local operations and classical communications. The local operations are a series of local quantum gates, and the classical communications are used for exchanging the measurement results in two nodes. The variational quantum algorithm shown in lower part of Fig. \ref{VQCEPP} (a) is a hybrid quantum-classical simulator using classical optimizer to optimize parameterized quantum circuits.
Here, we propose a general VQC framework shown in Fig. \ref{VQCEPP} (b) to simulate the entanglement purification with multi-DoF. Its main architecture includes ansatz, measurement and classical optimizer. In conventional entanglement purification protocols, local operations usually have two parts arranged before and after measurement, we assume the ansatz in our VQC framework expresses only the part before measurement for simplicity. The ansatz is the main concern in learning entanglement purification and prepared with parameterized quantum circuits based on practical conditions. The gate operations in the ansatz between circuit lines of Alice and Bob are forbidden because of only allowing local quantum operations. By optimizing the parameterized quantum circuits in the ansatz, the process of local operations in entanglement purification can be simulated effectively. Each circuit line represents one DoF of entangled particle.
We assume that Alice and Bob share $n$ pairs of entangled photons denoted by sets $A=\{A_{1}, A_{2},..., A_{n}\}$ and $B=\{B_{1}, B_{2},..., B_{n}\}$, respectively. $A_{i}$ and $B_{i}$ are entangled pair. The photons with $m$ DoFs in the sets A and B are also described as sets $A_{i}=\{D_{Ai1}, D_{Ai2},..., D_{Aim}\}$ and $B_{i}=\{D_{Bi1}, D_{Bi2},..., D_{Bim}\}$, and the elements $D_{Aij}$ and $D_{Bij}$ stand for the different DoFs of photons, such as polarization, space, and time. Preparing the quantum circuits in the VQC for entanglement purification, the circuit lines arranged for Alice and Bob's photons are represented with sets $C_{A}=\{C_{A1}, C_{A2}, ..., C_{An}\}$ and $C_{B}=\{C_{B1}, C_{B2}, ..., C_{Bn}\}$ respectively. All the elements in the sets $C_{Ai}$ and $C_{Bi}$ contain the circuit lines for the DoFs of photons and expressed by $C_{Ai}=\{L_{Ai1}, L_{Ai2}, ..., L_{Aim}\}$ and $C_{Bi}=\{L_{Bi1}, L_{Bi2}, ..., L_{Bim}\}$. The construction of the circuit lines in VQC is actually a map from entangled photons to lines, i.e., $f: D_{Aij} (D_{Bij})\mapsto L_{Aij} (L_{Bij})$. The dimension of each line $L_{Aij}$, $L_{Bij}$ is decided by $D_{Aij}$, $D_{Bij}$ and has the relation of $\text{dim}(L_{Aij})=\text{dim}(D_{Aij})$, $\text{dim}(L_{Bij})=\text{dim}(D_{Bij})$. An instance of photons in which two entangled pairs with two DoFs shared by Alice and Bob is given in Fig. \ref{VQCEPP} (b), the horizontal solid and dashed lines represent photonic polarization and spatial DoFs, respectively. The circuit lines $C_{A1}=\{L_{A11}, L_{A12}\}$, $C_{A2}=\{L_{A21}, L_{A22}\}$ belong to Alice and $C_{B1}=\{L_{B11}, L_{B12}\}$, $C_{B2}=\{L_{B21}, L_{B22}\}$ belong to Bob. The $C_{A1}$, $C_{B1}$ are nonlocal entangled pair, and the $C_{A2}$, $C_{B2}$ are another entangled one. The polarization has horizontal (H) and vertical (V) directions, and the spatial DoF is assumed with only upper and lower paths. Therefore, the dimension of DoF is $\text{dim}(L_{Aij})=\text{dim}(L_{Bij})=2$.
In entanglement purification, the goal is to obtain the output state with higher fidelity. The target state $|\psi_{target}\rangle$  usually is a pure state and the learning objective function is chosen with fidelity  $f_{out}=\langle\psi_{target}|\rho_{out}|\psi_{target}\rangle$ whose direction of optimization is maximization. Here, $\rho_{out}$ is the density operator of residual entangled pairs after entanglement purification, i.e. depending on $U(\bm{\theta})\rho_{in}U^{\dagger}(\bm{\theta})$ and postselection. Therefore, when the measurement outcome is chosen, the goal is to train the parameter vector $\bm{\theta}$. Figure \ref{VQCEPP} (c) shows the instances of the physical implementations of several basic quantum gates on polarization and spatial DoFs in linear optics. The half-wave plate can act as the Hadamard or Pauli-X gate of photonic polarization DoF with different input angles. The polarizing beam splitter (PBS) which reflects V and transmits H polarization of photon realizes a CNOT gate on photonic polarization (source) and spatial (target) DoFs. A reversed CNOT between polarization (target) and spatial (source) DoFs can be realized by adding a Pauli-X gate (a half-wave plate) in the lower path.

\section{Learning the entanglement purification with multi-DoF in linear optics}\label{secLEP}

\subsection{PSBZ protocol}\label{secpan}
\begin{figure}[]
\includegraphics[width=\linewidth]{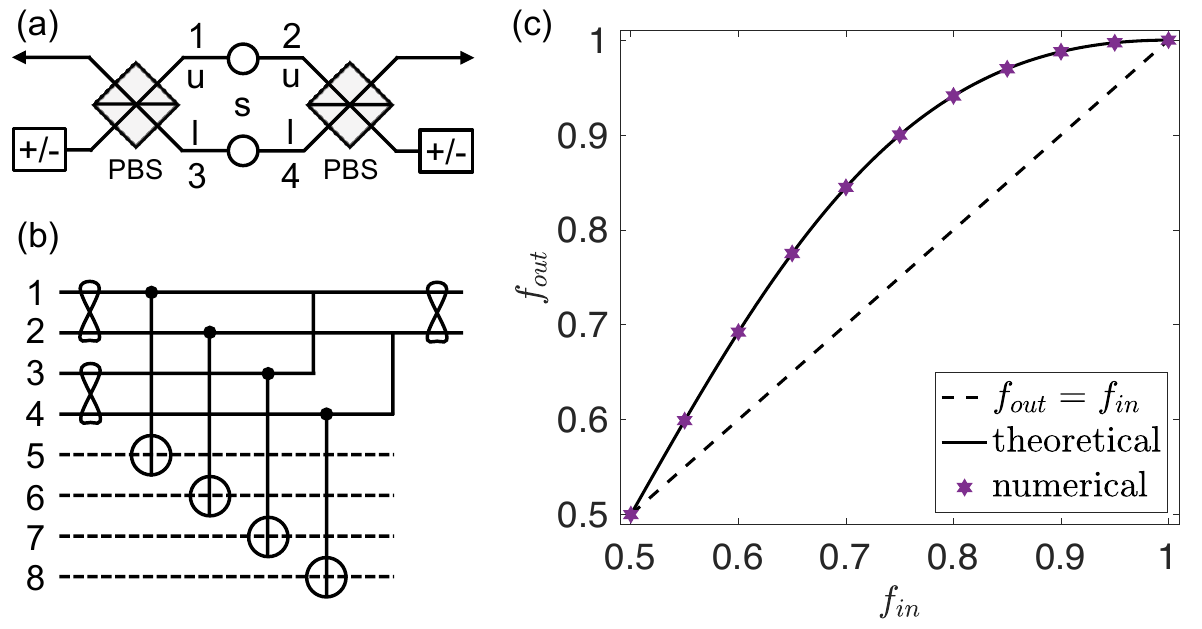}
\caption{The VQC learning of the PSBZ protocol. (a) The physical schematic diagrams of entanglement purification. ``s'' represents ideal entanglement source. ``u'' and ``l'' are upper and lower spatial DoF, respectively. (b) The quantum circuit version of PSBZ protocol. (c) The numerical and theoretical results of fidelities.}\label{panEPP}
\end{figure}

Four Bell states are considered in entanglement purification as follows
\begin{eqnarray}\nonumber
|\Phi^{\pm}\rangle&=&\frac{1}{\sqrt{2}}(|00\rangle\pm|11\rangle),\\
|\Psi^{\pm}\rangle&=&\frac{1}{\sqrt{2}}(|01\rangle\pm|10\rangle).\label{Bellstate}
\end{eqnarray}
To learn entanglement purification by our VQC framework, we first study  PSBZ protocol proposed by Pan \emph{et al.}  \cite{pannature1} for linear optics. In the protocol, photonic spatial DoF is introduced to overcoming the problem of difficult CNOT operations between photons. The ideal case is that Alice and Bob share Bell pairs $|\Phi^{+}_{ab}\rangle=\frac{1}{\sqrt{2}}(|0_{a}0_{b}\rangle\pm|1_{a}1_{b}\rangle)$ in polarization DoF from ideal source. Here, the states 0 and 1 in $|\Phi^{+}_{ab}\rangle$ represent for the V and H polarization of photon, respectively. The photons labeled with ``a'' and ``b'' belong to Alice and Bob, respectively. A mixed state considered with only bit-flip error before entanglement purification is given by
\begin{eqnarray}\label{panmixed}
\rho^{ab}_{in}=f_{in}|\Phi^{+}_{ab}\rangle\langle\Phi^{+}_{ab}|+(1-f_{in})|\Psi^{+}_{ab}\rangle\langle\Psi^{+}_{ab}|.
\end{eqnarray}
The two copies of this mixed state $\rho^{a1b1}_{in}\otimes\rho^{a2b2}_{in}$ have four components, including
$|\Phi^{+}_{a1b1}\rangle|\Phi^{+}_{a2b2}\rangle$ with fidelity $f_{in}^{2}$,
$|\Phi^{+}_{a1b1}\rangle|\Psi^{+}_{a2b2}\rangle$ with fidelity $f_{in}(1-f_{in})$,
$|\Psi^{+}_{a1b1}\rangle|\Phi^{+}_{a2b2}\rangle$ with fidelity $f_{in}(1-f_{in})$, and
$|\Psi^{+}_{a1b1}\rangle|\Psi^{+}_{a2b2}\rangle$ with fidelity $(1-f_{in})^{2}$.
Different with protocols \cite{BBPSSW} and \cite{DEJMPS}, the PSBZ protocol replaces the CNOT between two photons with the CNOT between the two DoFs of each photon. As shown in Fig. \ref{panEPP} (a), two symmetric PBSs are used by both Alice and Bob. The whole process is described with quantum circuit language in Fig. \ref{panEPP} (b). For simplicity of calculation, we label Alice and Bob's photons with 1, 3, 5, 7 and 2, 4, 6, 8, respectively. Circuit lines 1 (3) and 2 (4) are polarization entangled. All the circuit lines 5, 6, 7 and 8 are spatial DoF and plotted together. The initial state of spatial DoF is $|0_{5}0_{6}1_{7}1_{8}\rangle$ and 0 (1) stands for the upper (lower) path. The two PBSs play roles in applying four CNOT gates between the polarization and spatial DoFs of each photon. Actually, the CNOT gates on all spatial DoF produce the two pairs of four-qubit entangled state.  Alice and Bob should measure their lower spatial DoF with basis $|\pm\rangle=\frac{1}{\sqrt{2}}(|0\rangle\pm|1\rangle)$ and choose the case of all output with photon, i.e. so called `four-mode cases' \cite{pannature2}. 
The detailed derivation process of above protocol is given in Appendix \ref{apppsbz}.
If Alice and Bob get the result $|++\rangle$ or $|--\rangle$, the state of source pair is
\begin{eqnarray}\label{}
\rho^{\pm\pm}_{upper}=f^{\pm\pm}_{out}|\Phi^{+}\rangle\langle\Phi^{+}|+(1-f^{\pm\pm}_{out})|\Psi^{+}\rangle\langle\Psi^{+}|,
\end{eqnarray}
where the new fidelity is $f^{\pm\pm}_{out}=\frac{f_{in}^{2}}{f_{in}^{2}+(1-f_{in})^{2}}$.
When the result is $|+-\rangle$ or $|-+\rangle$, the output state will be translated to $|\Phi^{+}\rangle$, whose new fidelity is $f^{\pm\mp}_{out}=f^{\pm\pm}_{out}$, by applying a local phase flip gate on the one of the residual photons.

Using the VQC framework to learn the above entanglement purification protocol, we directly learn the ansatz by assuming a series of parameterized universal quantum gates including single qubit arbitrary rotation gates and two qubit CNOT gates. The input of ansatz is $\rho_{in}=\rho_{a1b1}\otimes\rho_{a2b2}\otimes|0_{5}0_{6}1_{7}1_{8}\rangle\langle 0_{5}0_{6}1_{7}1_{8}|$. Our goal is to optimize the fidelity of the final output state given by objective function
$f_{out}=\langle\Phi^{+}|\rho_{out}|\Phi^{+}\rangle$ and the $\rho_{out}$ is chosen with $\rho_{out}=\rho^{++}_{upper}$ in our simulations.
Numerical results are shown in Fig. \ref{panEPP} (c). The points in the figure are our numerical fidelities and match the theoretical curve $f^{\pm\pm}_{out}$ very well. Each point shown in the figure is the best one chosen from learning results by randomly initializing the ansatz with 10 times as the one time learning might give a local optimal fidelity rather than global optimal one. The learned ansatz suggests the optimal fidelities of entanglement purification are the same as the PSBZ protocol in Fig. \ref{panEPP} (b) for the input mixed state $\rho_{in}$. The learned local operations are not unique due to the learning of fidelity.

\begin{figure}[]
\includegraphics[width=\linewidth]{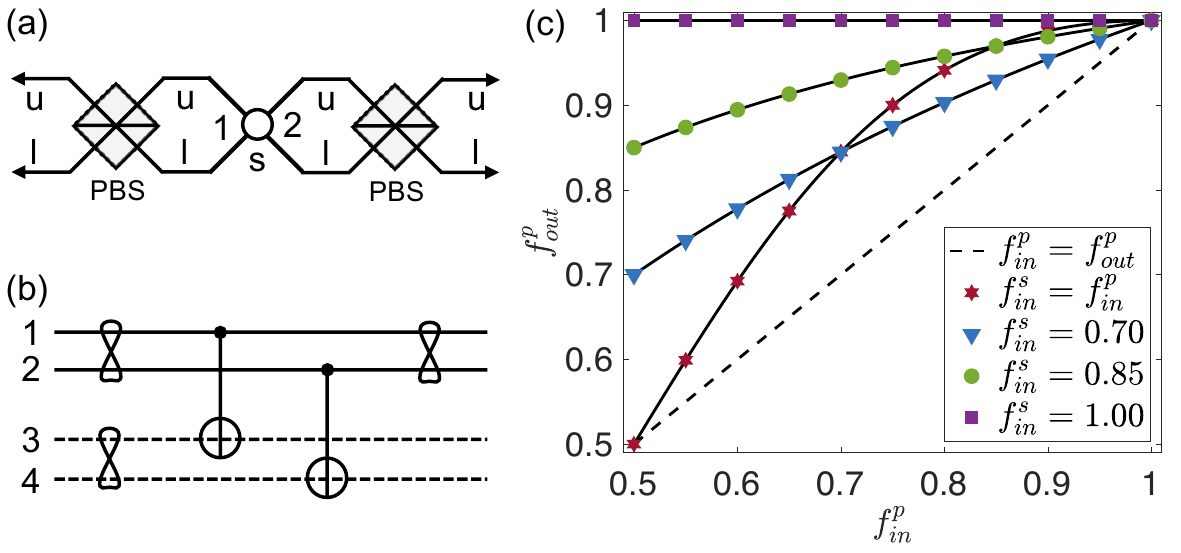}
\caption{The VQC learning of HHSZ+ and Simon-Pan protocols with hyperentanglement. (a) The physical schematic diagrams of the two protocols. (b) The quantum circuit version of the two protocols. Lines 1 (2) and 3 (4) are polarization and spatial DoFs of photon belonging to Alice (Bob), respectively. (c) The numerical and theoretical results of fidelities. The solid curves and markers are theoretical and numerical results, respectively.}\label{HHSZ}
\end{figure}

\subsection{Hyperentanglement-based protocols}\label{secsimonpan}
Another kind of entanglement purification protocols using multi-DoF is based on hyperentanglement. Only one pair of photons entangled in both polarization and spatial DoFs are required. The Bell states of polarization and spatial DoFs have the same form with Eq. (\ref{Bellstate}) and are respectively labeled with $|\Phi^{\pm}_{p}\rangle$, $|\Psi^{\pm}_{p}\rangle$, $|\Phi^{\pm}_{s}\rangle$ and $|\Psi^{\pm}_{s}\rangle$. Subscripts ``p'' and ``s'' represent polarization and spatial DoFs, respectively. The typical entanglement purification protocols based on hyperentanglement are HHSZ+ \cite{husheng}, Simon-Pan \cite{simonpan}, Li \cite{lionestep}, Sheng-Deng \cite{shengonestep} and etc. We first study the HHSZ+ protocol whose initial state is described by $\rho_{in}=\rho^{p}_{in}\otimes\rho^{s}_{in}$, where the density operators $\rho^{p}_{in}$ and $\rho^{s}_{in}$ are
\begin{eqnarray}\label{mixedp}
\rho_{in}^{p}&=&f_{in}^{p}|\Phi^{+}_{p}\rangle\langle\Phi^{+}_{p}|+(1-f_{in}^{p})|\Psi^{+}_{p}\rangle\langle\Psi^{+}_{p}|,
\end{eqnarray}
and
\begin{eqnarray}\label{mixeds}
\rho_{in}^{s}&=&f_{in}^{s}|\Phi^{+}_{s}\rangle\langle\Phi^{+}_{s}|+(1-f_{in}^{s})|\Psi^{+}_{s}\rangle\langle\Psi^{+}_{s}|.
\end{eqnarray}
Here, $f_{in}^{p}$ and $f_{in}^{s}$ are the fidelities of $|\Phi^{+}_{p}\rangle$ and $|\Phi^{+}_{s}\rangle$, respectively. The equivalent entanglement purification using time DoF is protocol \cite{ursinprl}.
Here, only one entangled pair and two PBSs are required for accomplishing the entanglement purification of the polarization DoF shown in Fig. \ref{HHSZ} (a) when the input fidelities satisfy $f_{in}^{p}>\frac{1}{2}$ and $f_{in}^{s}>\frac{1}{2}$. In the VQC framework, two polarization and another two spatial lines are introduced in Fig. \ref{HHSZ} (b). Compared with the PSBZ protocol, the physical devices are the same but the quantum circuit only has two CNOT gates in the HHSZ+ protocol. Shown in Appendix \ref{secHHSZ}, the output fidelity is $f^{p}_{out}=\frac{f_{in}^{p}f_{in}^{s}}{f_{in}^{p}f_{in}^{s}+(1-f_{in}^{p})(1-f_{in}^{s})}$ for the selection of $|0_{3}0_{4}\rangle$ (two upper) or $|1_{3}1_{4}\rangle$ (two lower). Three cases are considered as $f_{in}^{p}=f_{in}^{s}$, fixed $f_{in}^{s}\in(0.5,1.0)$ and $f_{in}^{s}=1$ in numerical simulations. For the fixed $f_{in}^{s}\in(0.5,1.0)$, without loss of generality, we learn the output fidelities by fixing the $f_{in}^{s}$ with 0.70 and 0.85. When the fidelity satisfies $f_{in}^{s}=1$, HHSZ+ protocol transforms to the Simon-Pan protocol. As shown in Fig. \ref{HHSZ} (c), the curve of $f_{in}^{p}=f_{in}^{s}$ is the same with PSBZ protocol. Each point shown in the figure is also the best one chosen from learning results by random initializations with 10 times for avoiding local optimum. For instance, the case with input fidelities $f_{in}^{s}=0.70$ and $f_{in}^{p}=0.55$ is executed with given ansatz, the optimal output fidelity is $f_{out}^{p}=0.7404$. But there exist a local optimal point $f_{out}^{p}=0.70$ which suggests a swap operation to exchange the states between polarization and spatial DoFs. This operation will obtain a higher fidelity in polarization DoF, i.e. $f_{out}^{p}=0.70>f_{in}^{p}=0.55$, but is not global optimum. In $f_{in}^{s}=1$, the output purified state is a pure entangled state. Results in Fig. \ref{HHSZ} (c) indicate all the assumed cases are discovered well by our VQC framework.

\begin{figure}[]
\includegraphics[width=\linewidth]{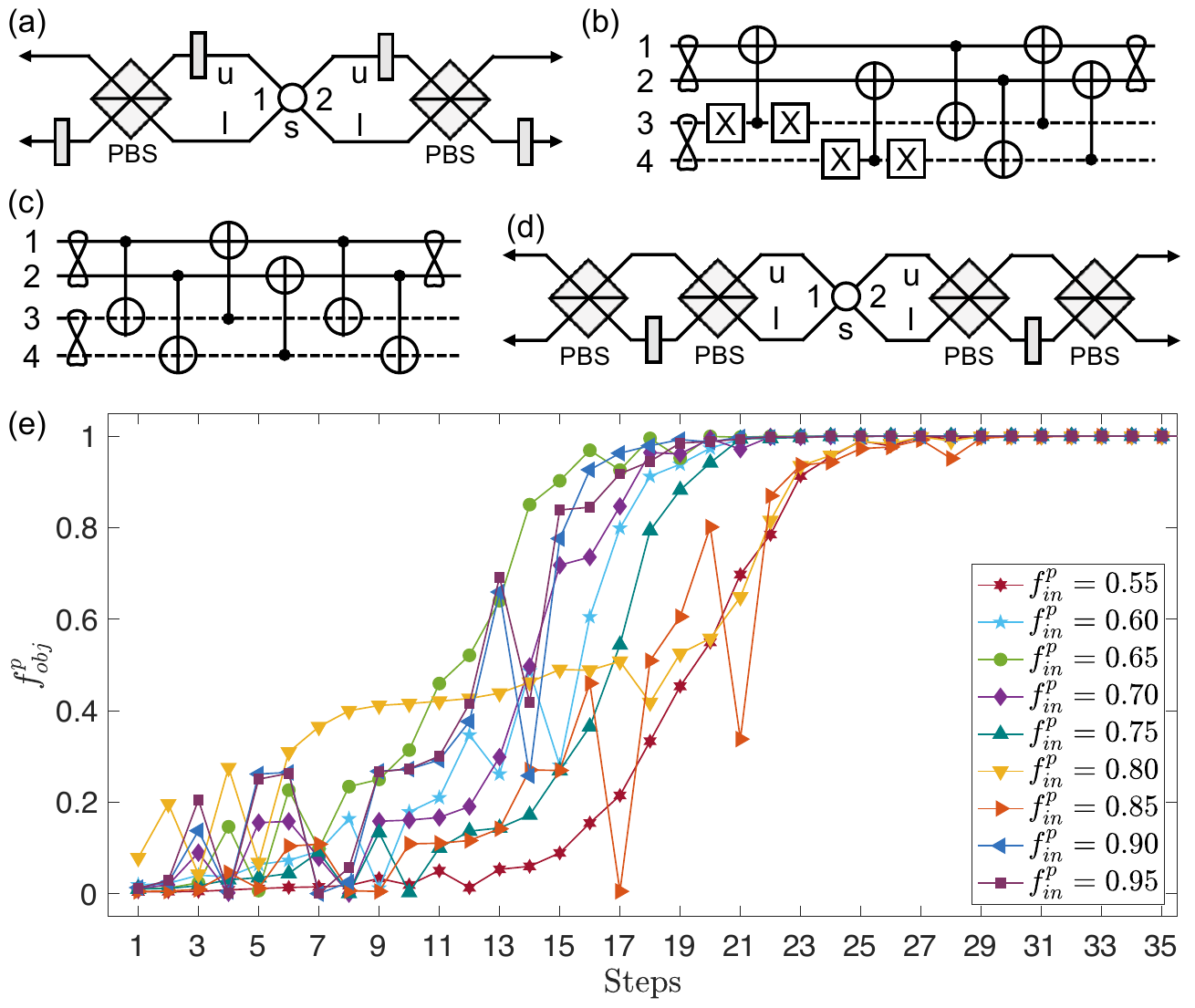}
\caption{The VQC learning of the Li protocol with hyperentanglement. (a) The physical schematic diagrams of the Li protocol. (b) The corresponding quantum circuit of the Li protocol. Lines 1 (2) and 3 (4) are polarization and spatial DoFs of photon belonging to Alice (Bob), respectively. (c) The quantum circuit of an example of new equivalent entanglement purification protocols learned by VQC. (d) The physical schematic diagrams of the new equivalent entanglement purification circuit in (c). (e) The learning curves of chosen fidelities based on VQC with a set of random initial parameters.}\label{EPPLi}
\end{figure}

If we study the case considering phase errors, terms $|\Phi^{-}\rangle\langle\Phi^{-}|$ and $|\Psi^{-}\rangle\langle\Psi^{-}|$ are added in density operators. When spatial DoF is a pure entangled state, the corresponding protocols are the Li \cite{lionestep} and Sheng-Deng \cite{shengonestep} protocols. As an instance, we simulate Li protocol here. In Li's protocol, all the output states appearing in spatial DoF $|00\rangle$,  $|01\rangle$, $|10\rangle$ and $|11\rangle$ are $|\Phi^{+}_{p}\rangle$. To keep the same target with Li's protocol, we use a new objective function with $f_{obj}^{p}=f_{00}^{p}* f_{01}^{p}*f_{10}^{p}* f_{11}^{p}$, where $f_{00}^{p}$, $f_{01}^{p}$, $f_{10}^{p}$ and $f_{11}^{p}$ are the fidelities of output states in the spatial DoF $|00\rangle$ (two upper), $|01\rangle$(one upper one lower), $|10\rangle$ (one upper one lower) and $|11\rangle$ (two lower), respectively. One can find that when the objective function $f_{obj}^{p}$ is optimized to 1, all the four fidelities $f_{00}^{p}$, $f_{01}^{p}$, $f_{10}^{p}$, $f_{11}^{p}$ are 1. The detailed explanations of the Li protocol are given in Appendix \ref{secLi}. Compared with the Simon-Pan protocol, the operations in this case need three bilateral CNOT gates in Alice and Bob. The physical implementation and its corresponding quantum circuit of the Li protocol are plotted in Figs. \ref{EPPLi} (a) and (b), respectively. By the VQC learning, Li protocol is also learned well. Moreover, the results of optimized parameters in the ansatz also suggest some other different local operations whose final outputs are the same with those of Li's. We show one of the equivalent protocols discovered by the VQC learning in Fig. \ref{EPPLi} (c) and (d) with quantum circuit and setup versions where four PBSs and two half-wave plates are used. This simple result can be learned by adaptively reducing and adjusting the gate operations in the ansatz.  The learning curves of fidelities chosen with one of the best results from randomly initializations 10 times for each point are shown in Fig. \ref{EPPLi} (e). Here, the initial input fidelities of each states are assumed with $f^{p}_{in,\Phi^{-}}=(1-f^{p}_{in,\Phi^{+}})/2$, $f^{p}_{in,\Psi^{+}}=(1-f^{p}_{in,\Phi^{+}})/4$, $f^{p}_{in,\Psi^{-}}=(1-f^{p}_{in,\Phi^{+}})/4$ and $f^{s}_{in,\Phi^{+}}=1$. With about 30 steps (actual steps not shown in the figure, breaking up the optimizing processes is usually more than 35 steps), all the 9 learning processes, i.e. input with $f^{p}_{in,\Phi^{+}}=\{0.55,0.60,...,0.90,0.95\}$, achieve their optimal values.

\section{Entanglement purification of multi-pair and multi-DoF input}\label{secscalability}

\begin{figure}[]
\centering
\includegraphics[width=\linewidth]{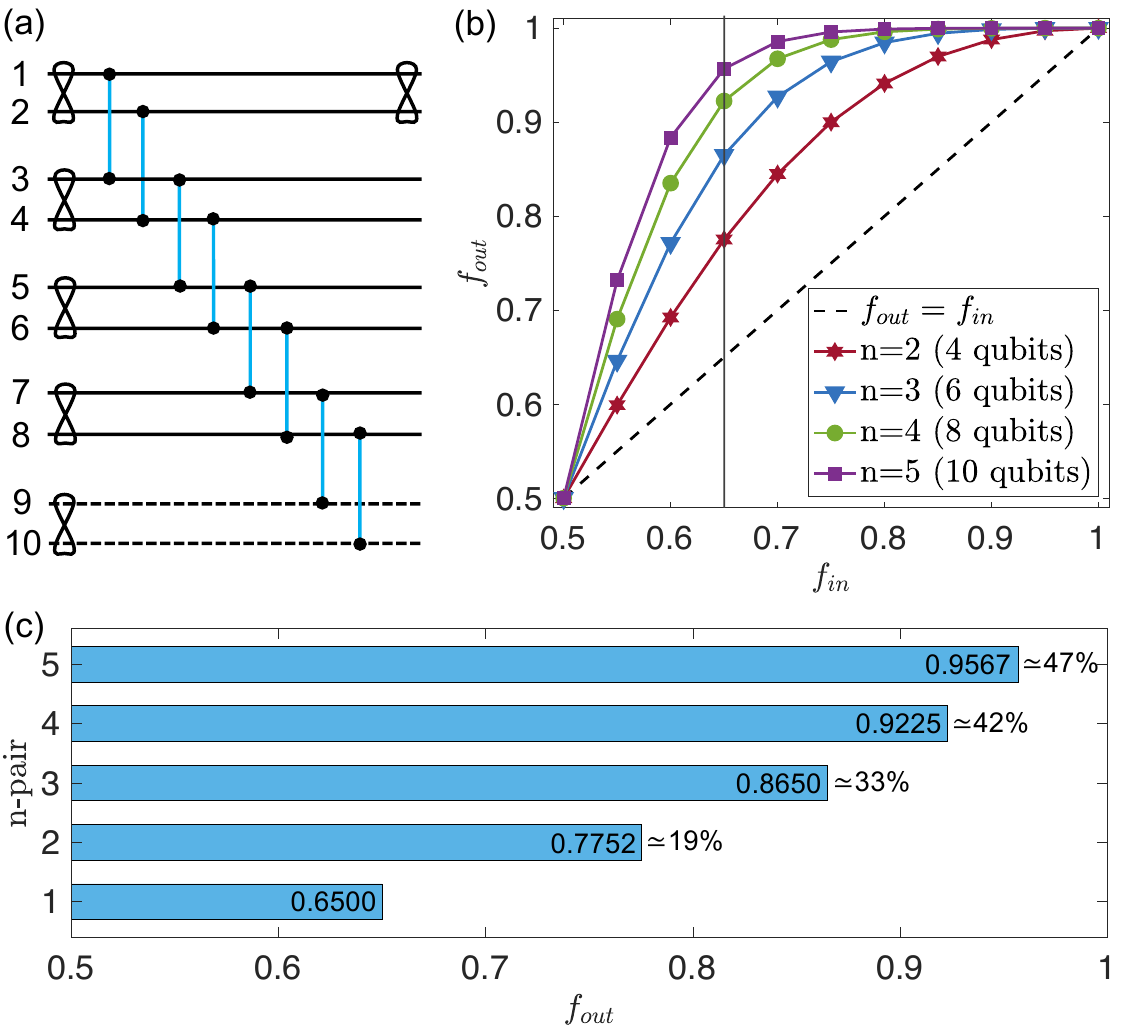}
\caption{Entanglement purification for the cases of multi-pair with multi-DoF. (a) The purification model and ansatz structure of five entangled pairs (10 qubits). The connection labeled with a vertical line and two dots between two circuit lines is two-qubits ansatz. (b) The learning fidelities of all cases from two to five input entangled pairs. (c) The learning fidelities of $f_{in}=0.65$ from two to five input entangled pairs.}\label{epmultipair}
\end{figure}

To verify the scalability of the framework, we simulate the entanglement purification of mutipair cases with multi-DoF and show the results of two pair (4 qubits), three pair (6 qubits), four pair (8 qubits), and  five pair (10 qubits). As shown in Fig. \ref{epmultipair} (a), the model given by the case of five entangled pairs (10 qubits) is chosen with applying the two-qubit ansatz between nearest-neighbor qubits for simplicity, and one ancillary DoF is considered. The initial input entangled states are assumed as
\begin{eqnarray}\label{multipair}
\rho_{in}=\bigotimes^{n}_{i=1}\rho^{i}_{in},
\end{eqnarray}
where $n$ is considered with $n= 2, 3, 4, 5$ corresponding to $4, 6, 8, 10$ qubits, and for simplicity, all the $\rho^{i}_{in}$ have same fidelities as
\begin{eqnarray}\label{multiinput}
\rho_{in}^{i}&=&f_{in}|\Phi^{+}_{i}\rangle\langle\Phi^{+}_{i}|+(1-f_{in})|\Psi^{+}_{i}\rangle\langle\Psi^{+}_{i}|.
\end{eqnarray}
We assume that the residual pair is kept when we chose the case of $|0\rangle^{\otimes2(n-1)}$ or $|1\rangle^{\otimes2(n-1)}$ of the $n-1$ nonlocal entangled pairs. The output density matrix calculated in objective function is chosen with $\rho_{out}=\rho^{1}_{out}+\rho^{0}_{out}$, where $\rho^{1}_{out}$ ($\rho^{0}_{out}$) is residual state corresponding to measurement outcome $|0\rangle^{\otimes2(n-1)}$ ($|1\rangle^{\otimes2(n-1)}$). In numerical simulations, input fidelities $\{0.5, 0.55, 0.6, ..., 0.95, 1.0\}$ are performed. For the each point of the input fidelities, we also simulate it for 10 times by initializing the parameters of ansatz randomly and choose the best fidelity as the final output result rather than expectation. For the case of five pair (10 qubits), when the output optimal numerical result is larger than the optimal fidelity of the 4-pair case, the simulation is ended at this round for saving time. All the purification results shown in Fig. \ref{epmultipair} (b) indicate that the output fidelity can be increased by using more input pairs in domain $f_{in}\in(0.5,1.0)$. For instance of $f_{in}=0.65$ shown in Fig. \ref{epmultipair} (c), the fidelity is purified to $f_{out}\simeq0.7752$ (purification rate $\text{pr}=(f_{out}-f_{in})/f_{in}\simeq19\%$) in the two-pair case and this result can be further enhanced to $f_{out}\simeq0.8650$ ($\text{pr}\simeq33\%$), $f_{out}\simeq0.9225$ ($\text{pr}\simeq42\%$) and $f_{out}\simeq0.9567$ ($\text{pr}\simeq47\%$) in the cases of three pair, four pair and five pair, respectively. Our calculations indicate that the entanglement purification with multi-pair cases can be carried out with only one round.
In conventional protocols, the purifications of multi-pair mixed entangled states are usually based on a recursive process in which 2-pair case is carried out in each round. For the instance of four entangled pairs' input, the conventional recursive process is two rounds described by $\text{4 pairs}\rightarrow\text{2 pairs}\rightarrow\text{1 pairs}$, but this process is only one round in our protocol, i.e. $\text{4 pairs}\rightarrow\text{1 pairs}$.

\begin{figure}[]
\centering
\includegraphics[width=\linewidth]{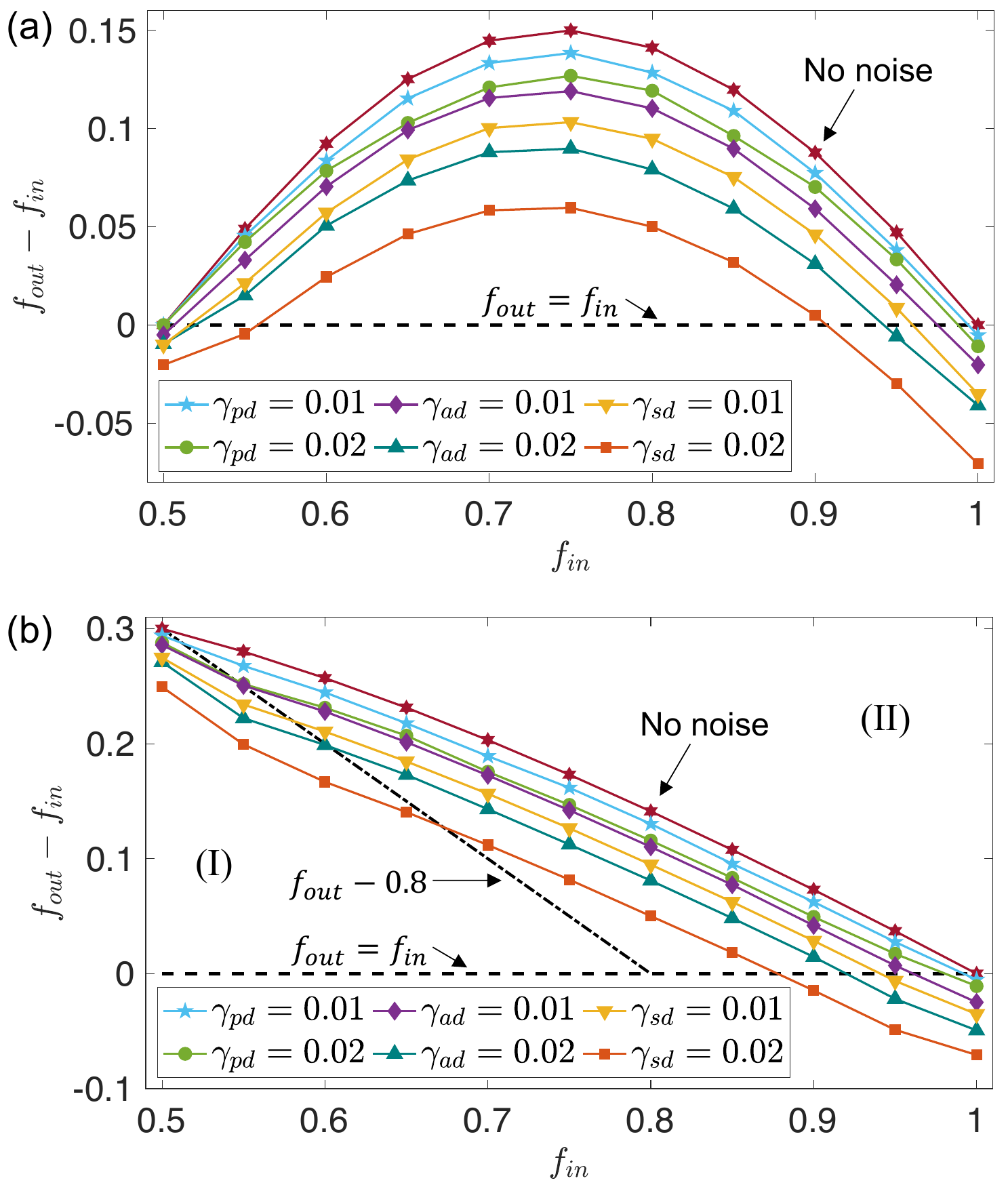}
\caption{The simulation of entanglement purification with noise. (a) The results of the case with the input entangled states $f_{in}=f_{in}^{i}=f_{in}^{j}$. (b) The results of the case with the input entangled states $f_{in}=f_{in}^{i}$ and $f_{in}^{j}=0.80$. }\label{epnoise}
\end{figure}

\section{Entanglement purification with noise}\label{secnoise}

To simulate practical cases, we consider the three types of noisy channels, i.e. phase damping, amplitude damping and symmetric depolarizing channel, in our simulations of entanglement purification. The noisy channels are described by applying Kraus operators $E_{k}$ to density matrix by conjugation as follows,
\begin{eqnarray}\label{Kraus}
\rho=\mathbb{N}(\rho)=\sum_{k}E_{k}\rho E_{k}^{\dag}.
\end{eqnarray}
The Kraus operators of phase damping channel $\mathbb{N}_{\text{pd}}$ are given by
\begin{eqnarray}\label{phasedamping}
E_{1}=\left[
        \begin{array}{cc}
          1 & 0 \\
          0 & \sqrt{1-\gamma} \\
        \end{array}
      \right],\
E_{2}=\left[
        \begin{array}{cc}
          0 & 0 \\
          0 & \sqrt{\gamma} \\
        \end{array}
      \right].
\end{eqnarray}
For amplitude damping channel $\mathbb{N}_{\text{ad}}$, the Kraus operators has form of
\begin{eqnarray}\label{amplitudedamping}
E_{1}=\left[
        \begin{array}{cc}
          1 & 0 \\
          0 & \sqrt{1-\gamma} \\
        \end{array}
      \right],\
E_{2}=\left[
        \begin{array}{cc}
          0 & \sqrt{\gamma} \\
          0 & 0 \\
        \end{array}
      \right].
\end{eqnarray}
Symmetric depolarizing channel is described by a set of Pauli operators as
\begin{eqnarray}\label{depolarizing}
\mathbb{N}_{\text{sd}}(\rho)=(1-\frac{3\gamma}{4})\rho+\frac{\gamma}{4}(\sigma_x\rho\sigma_x+\sigma_y\rho\sigma_y+\sigma_z\rho\sigma_z).
\end{eqnarray}
The parameter $\gamma$ in above expressions stands for the magnitude of the noise.

Two-qubit gates have higher noise rate than one-qubit ones in physical considerations \cite{Enoise}; for simplicity, the noise is applied only after CNOT gates in our simulations. The two cases of initial input fidelities, i.e., $f_{in}^{i}=f_{in}^{j}$ and $f_{in}=f_{in}^{i}$ with fixed $f_{in}^{j}=0.80$, are simulated by considering noisy channel with $\gamma=0, 0.01, 0.02$.  The learning results of entanglement purification with noise are shown in Figs. \ref{epnoise} (a) and (b). The measurement choices of all types of noise channel are the same, i.e., choosing $|00\rangle$ or $|11\rangle$ of ancillary DoF, and each data point shown in the figure is the best result of random initializations with 10 times. The objective function is chosen with fidelity $f_{out}=\langle\Phi^{+}|(\rho^{00}_{out}+\rho^{11}_{out})|\Phi^{+}\rangle$, where $\rho^{00}_{out}$ ($\rho^{11}_{out}$) is the output state of measurement outcome 00 (11). $\gamma_{pd}$, $\gamma_{ad}$ and $\gamma_{sd}$ represent the magnitude of the noise of phase damping, amplitude damping and symmetric depolarizing channel, respectively. In the case of $f_{in}^{i}=f_{in}^{j}$, we calculate the result of $f_{out}-f_{in}$ called net fidelity. While in the case of fixed $f_{in}^{j}=0.8$, we choose the net fidelity $f_{out}-f_{in}$ in which $f_{in}$ is chosen with the modulated input fidelity $f_{in}^{i}$ rather than the fixed one $f_{in}^{j}$. The results in the two figures show that noise reduces the quality of entanglement purification. As the noise ($\gamma$) increases, the net fidelities decrease. If the net fidelity is less than zero where is the area below dashed line $f_{out}=f_{in}$ in Fig. \ref{epnoise} (a), it indicates the purification fails. In Fig. \ref{epnoise} (b), the case is little different, we give two dashed lines, i.e., the line $f_{out}=f_{in}$ and $f_{out}-0.8$. In  some cases of multi-DoF purification, for instance in linear optics, one usually uses ancillary DoF, i.e., path, to purify the primary one, i.e., polarization. If the output fidelity of polarization is greater than the input, the purification is valid although the output fidelity is less than the input fidelity of spatial DoF. We give a dashed line $f_{out}=f_{in}$ as the critical line for the above case and a line $f_{out}-0.8$ for a common case whose requirement is $f_{out}$ greater than all the input fidelities. So in our simulations, the second dashed line is $f_{out}-0.8$ as the fidelity of input ancillary DoF is 0.8. In Fig. \ref{epnoise} (b), we mark two valid areas $\mathrm{I}+\mathrm{II}$ and $\mathrm{II}$ for the requirements of $f_{out}>f_{in}$ and $f_{out}> f_{in}\  \&\  0.8$, respectively.  Comparing three types of noise, the phase-damping and symmetric depolarizing channels have the smallest and biggest influence on entanglement purification, respectively.

\section{Discussion and Summary}\label{secDS}
\begin{figure}[]
\centering
\includegraphics[width=\linewidth]{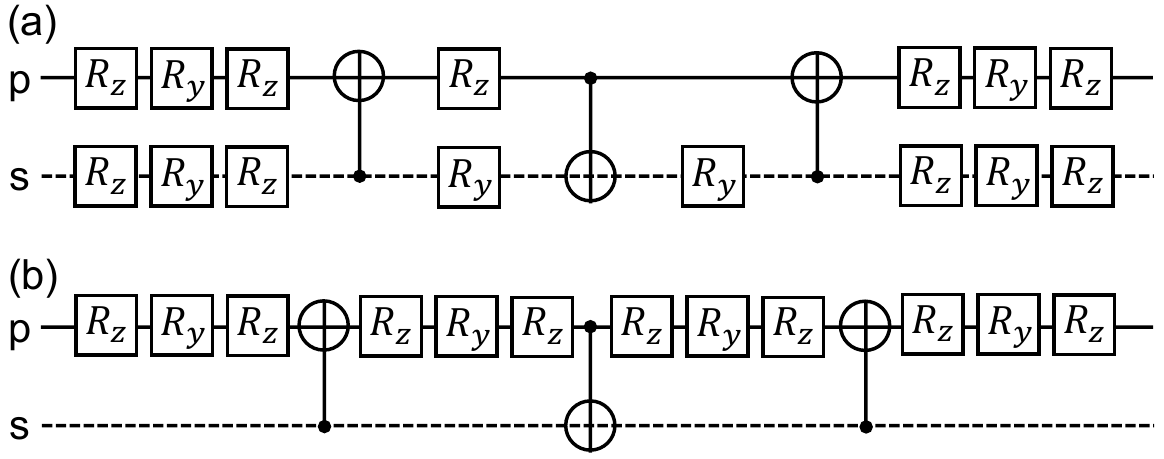}
\caption{The quantum circuits of the ansatz used in our simulations between two different DoFs. (a) An universal quantum circuit for two-qubit operations. (b) A limited ansatz with no single-qubit operations in spatial DoF. $R_{y}$ and $R_{z}$ are single-qubit rotation gates about the y and z axes, respectively. }\label{ansatz}
\end{figure}

The circuit of ansatz with parameters $\bm{\theta}_u=[\theta_1, \theta_2, ... , \theta_{15}]$ used in our simulations is given in Fig. \ref{ansatz} (a). It is an universal quantum circuit for two-qubit operations with a series of single-qubit operations and three CNOT gates \cite{gateVQC1,gateVQC2}. In some practical cases, the operations can not be realized with experimental conditions. Therefore, we introduce a limited ansatz with $\bm{\theta}_l=[\theta_1, \theta_2, ... , \theta_{12}]$ by assuming the single qubit operations are difficult in the spatial DoF in Fig. \ref{ansatz} (b). In practical numerical simulations, we learned the optimal fidelities by using a universal ansatz and also can repeat the process with a limited one according to experimental conditions if necessary. The parameter vector $\bm{\theta}_u$ in the ansatz is assigned with random numbers in domain $[0,2\pi]$, so the learning curves are usually different in each round, but the final stationary results of fidelities are almost the same (local optimum can be detoured by repeating learning process more times).  Our numerical simulations are mainly performed by using MINDQUANTUM \cite{codes}. The optimizer is chosen with the Broyden-Fletcher-Goldfarb-Shanno (BFGS) \cite{BFGS1,BFGS2,BFGS3,BFGS4}.

In conclusion, we have proposed an effective VQC framework for learning the entanglement purification with multi-DoF by properly introducing  additional circuit lines. To verify our framework, the well-known protocols of entanglement purification in linear optics, such as PSBZ, HHSZ+, Simon-Pan, etc., are learned with alternative operations and the fidelities match well with the theoretical values. To show the scalability, we simulate the cases of multi-pair with multi-DoF well and show that the entanglement purification of multi-pair can be carried out with only one round. Moreover, the influence of noise is performed. The results indicate that our VQC learning method is effective in designing the protocols of the entanglement purification in multi-DoF. Our work introduces an effective way to design entanglement purification by variational quantum learning with near-term quantum devices.



\appendix
\section{PSBZ protocol in quantum circuit language}\label{apppsbz}
The density operator of an initial mixed state is described as
\begin{eqnarray}\label{}
\rho_{12}=\rho_{34}=f_{in}|\Phi^{+}_{ab}\rangle\langle\Phi^{+}_{ab}|+(1-f_{in})|\Psi^{+}_{ab}\rangle\langle\Psi^{+}_{ab}|.
\end{eqnarray}
Therefore, the system composed of two entangled pairs has density operator given by
\begin{eqnarray}\label{}\nonumber
\rho_{1-4}&=&\rho_{12}\otimes\rho_{34}\\\nonumber
&=&[f_{in}^{2}|\Phi^{+}_{12}\rangle\langle\Phi^{+}_{12}|\otimes|\Phi^{+}_{34}\rangle\langle\Phi^{+}_{34}|\\\nonumber
&&+f_{in}(1-f_{in})|\Phi^{+}_{12}\rangle\langle\Phi^{+}_{12}|\otimes|\Psi^{+}_{34}\rangle\langle\Psi^{+}_{34}|\\\nonumber
&&+f_{in}(1-f_{in})|\Psi^{+}_{12}\rangle\langle\Psi^{+}_{12}|\otimes|\Phi^{+}_{34}\rangle\langle\Phi^{+}_{34}|\\
&&+(1-f_{in})^{2}|\Psi^{+}_{12}\rangle\langle\Psi^{+}_{12}|\otimes|\Psi^{+}_{34}\rangle\langle\Psi^{+}_{34}|].
\end{eqnarray}
The state of spatial DoF at the beginning is a product state expressed with
$\rho_{5-8}=|0_{5}0_{6}1_{7}1_{8}\rangle\langle 0_{5}0_{6}1_{7}1_{8}|$,
where 0 and 1 stand for upper and lower paths, respectively. The density operator of the whole system can be written as
\begin{eqnarray}\label{}\nonumber
\rho_{1-8}&=&\rho_{1-4}\otimes\rho_{5-8}\\\nonumber
&=&f_{1}(|0_{1}0_{2}0_{3}0_{4}0_{5}0_{6}1_{7}1_{8}\rangle+|1_{1}1_{2}0_{3}0_{4}0_{5}0_{6}1_{7}1_{8}\rangle\\\nonumber
&&+|0_{1}0_{2}1_{3}1_{4}0_{5}0_{6}1_{7}1_{8}\rangle+|1_{1}1_{2}1_{3}1_{4}0_{5}0_{6}1_{7}1_{8}\rangle)\langle...|\\\nonumber
&&+f_{2}(|0_{1}0_{2}0_{3}1_{4}0_{5}0_{6}1_{7}1_{8}\rangle+|1_{1}1_{2}0_{3}1_{4}0_{5}0_{6}1_{7}1_{8}\rangle\\\nonumber
&&+|0_{1}0_{2}1_{3}0_{4}0_{5}0_{6}1_{7}1_{8}\rangle+|1_{1}1_{2}1_{3}0_{4}0_{5}0_{6}1_{7}1_{8}\rangle)\langle...|\\\nonumber
&&+f_{2}(|0_{1}1_{2}0_{3}0_{4}0_{5}0_{6}1_{7}1_{8}\rangle+|1_{1}0_{2}0_{3}0_{4}0_{5}0_{6}1_{7}1_{8}\rangle\\\nonumber
&&+|0_{1}1_{2}1_{3}1_{4}0_{5}0_{6}1_{7}1_{8}\rangle+|1_{1}0_{2}1_{3}1_{4}0_{5}0_{6}1_{7}1_{8}\rangle)\langle...|\\\nonumber
&&+f_{3}(|0_{1}1_{2}0_{3}1_{4}0_{5}0_{6}1_{7}1_{8}\rangle+|1_{1}0_{2}0_{3}1_{4}0_{5}0_{6}1_{7}1_{8}\rangle\\\nonumber
&&+|0_{1}1_{2}1_{3}0_{4}0_{5}0_{6}1_{7}1_{8}\rangle+|1_{1}0_{2}1_{3}0_{4}0_{5}0_{6}1_{7}1_{8}\rangle)\langle...|.
\end{eqnarray}
Here, the symbols $\langle...|$ mean they are the bras of corresponding kets in their left brackets. The fidelities $f_{1}$, $f_{2}$ and $f_{3}$ are  $f_{1}=f_{in}^{2}$, $f_{2}=f_{in}(1-f_{in})$ and $f_{3}=(1-f_{in})^{2}$.
PBS has the function of a CNOT gate whose control qubit is polarization and the target is spatial DoF as follows
\begin{eqnarray}\label{CNOT}\nonumber
U^{PBS}_{CNOT}&=&\bigotimes^{4}_{i=1}(|0_{i}0_{i+4}\rangle\langle 0_{i}0_{i+4}|+|0_{i}1_{i+4}\rangle\langle 0_{i}1_{i+4}|\\
&&+|1_{i}1_{i+4}\rangle\langle 1_{i}0_{i+4}|+|1_{i}0_{i+4}\rangle\langle 1_{i}1_{i+4}|).
\end{eqnarray}
Here, the states 0 and 1 represent V and H polarization, respectively.
After the CNOT gate operations on the states of all the spatial DoF, the two entangled pairs become
\begin{eqnarray}\label{}
\rho^{cnot}_{1-8}&=&U^{PBS}_{CNOT}\rho_{1-8}U^{PBS\dagger}_{CNOT}\\\nonumber
&=&f_{1}(|0_{1}0_{2}0_{3}0_{4}0_{5}0_{6}1_{7}1_{8}\rangle+|1_{1}1_{2}0_{3}0_{4}1_{5}1_{6}1_{7}1_{8}\rangle\\\nonumber
&&+|0_{1}0_{2}1_{3}1_{4}0_{5}0_{6}0_{7}0_{8}\rangle+|1_{1}1_{2}1_{3}1_{4}1_{5}1_{6}0_{7}0_{8}\rangle)\langle...|\\\nonumber
&&+f_{2}(|0_{1}0_{2}0_{3}1_{4}0_{5}0_{6}1_{7}0_{8}\rangle+|1_{1}1_{2}0_{3}1_{4}1_{5}1_{6}1_{7}0_{8}\rangle\\\nonumber
&&+|0_{1}0_{2}1_{3}0_{4}0_{5}0_{6}0_{7}1_{8}\rangle+|1_{1}1_{2}1_{3}0_{4}1_{5}1_{6}0_{7}1_{8}\rangle)\langle...|\\\nonumber
&&+f_{2}(|0_{1}1_{2}0_{3}0_{4}0_{5}1_{6}1_{7}1_{8}\rangle+|1_{1}0_{2}0_{3}0_{4}1_{5}0_{6}1_{7}1_{8}\rangle\\\nonumber
&&+|0_{1}1_{2}1_{3}1_{4}0_{5}1_{6}0_{7}0_{8}\rangle+|1_{1}0_{2}1_{3}1_{4}1_{5}0_{6}0_{7}0_{8}\rangle)\langle...|\\\nonumber
&&+f_{3}(|0_{1}1_{2}0_{3}1_{4}0_{5}1_{6}1_{7}0_{8}\rangle+|1_{1}0_{2}0_{3}1_{4}1_{5}0_{6}1_{7}0_{8}\rangle\\\nonumber
&&+|0_{1}1_{2}1_{3}0_{4}0_{5}1_{6}0_{7}1_{8}\rangle+|1_{1}0_{2}1_{3}0_{4}1_{5}0_{6}0_{7}1_{8}\rangle)\langle...|.
\end{eqnarray}
The four states of spatial DoF 
$|0_{5}0_{6}1_{7}1_{8}\rangle$, $|1_{5}1_{6}0_{7}0_{8}\rangle$, $|0_{5}1_{6}1_{7}0_{8}\rangle$, $|1_{5}0_{6}0_{7}1_{8}\rangle$ stand for the cases where each output port has one photon, i.e., so-called ``four-mode cases'' \cite{pannature2}. When we make a postselection of the above case, the states are projected into 
\begin{eqnarray}\label{}\nonumber
\rho^{cnot}_{1-8}
&=&f_{1}(|0_{1}0_{2}0_{3}0_{4}0_{5}0_{6}1_{7}1_{8}\rangle
\!+\!|1_{1}1_{2}1_{3}1_{4}1_{5}1_{6}0_{7}0_{8}\rangle)\langle...|\\\nonumber
&&\!+\!f_{3}(|0_{1}1_{2}0_{3}1_{4}0_{5}1_{6}1_{7}0_{8}\rangle
\!+\!|1_{1}0_{2}1_{3}0_{4}1_{5}0_{6}0_{7}1_{8}\rangle)\langle...|.
\end{eqnarray}
For the convenience of analysis, we rewrite the state in order from left upper and lower to right upper and lower of the spatial DoF, i.e., $|0_{5}1_{6}0_{7}1_{8}\rangle$. Under the new order, the above density matrix becomes
\begin{eqnarray}\label{rho18}\nonumber
\rho^{cnot}_{1-8}
&=&f_{1}(|0_{1}0_{2}0_{3}0_{4}0_{5}1_{6}0_{7}1_{8}\rangle
\!+\!|1_{1}1_{2}1_{3}1_{4}0_{5}1_{6}0_{7}1_{8}\rangle)\langle...|\\\nonumber
&&\!+\!f_{3}(|0_{1}0_{2}1_{3}1_{4}0_{5}1_{6}0_{7}1_{8}\rangle
\!+\!|1_{1}1_{2}0_{3}0_{4}0_{5}1_{6}0_{7}1_{8}\rangle)\langle...|.
\end{eqnarray}
In the PSBZ protocol, this measurement is implemented by measuring the photons in two lower paths with bases $|+\rangle=\frac{1}{\sqrt{2}}(|0\rangle+|1\rangle)$ and $|-\rangle=\frac{1}{\sqrt{2}}(|0\rangle-|1\rangle)$. Therefore, when the measurement result in Alice and Bob is $|++\rangle$ or $|--\rangle$, the state is chosen with
\begin{eqnarray}\label{}\nonumber
\rho^{\pm\pm}_{upper}&=&\rho^{++}_{upper}=\rho^{--}_{upper}\\\nonumber
&=&f_{1}(|0_{1}0_{3}\rangle+|1_{1}1_{3}\rangle)\langle...|\otimes|0_{5}0_{7}\rangle\langle...|\\
&&+f_{3}(|0_{1}1_{3}\rangle+|1_{1}0_{3}\rangle)\langle...|\otimes|0_{5}0_{7}\rangle\langle...|.\\\nonumber
\end{eqnarray}
Omitting the spatial DoF, the residual state of upper path is given by
\begin{eqnarray}\label{}\nonumber
\rho^{\pm\pm}_{upper}=f_{in}^{2}|\Phi^{+}\rangle\langle\Phi^{+}|+(1-f_{in})^{2}|\Psi^{+}\rangle\langle\Psi^{+}|.
\end{eqnarray}
If measurement result is $|+-\rangle$ or $|-+\rangle$, the state got by Alice and Bob is
\begin{eqnarray}\label{}\nonumber
\rho^{\pm\mp}_{upper}&=&\rho^{+-}_{upper}=\rho^{-+}_{upper}\\\nonumber
&=&f_{in}^{2}|\Phi^{-}\rangle\langle\Phi^{-}|+(1-f_{in})^{2}|\Psi^{-}\rangle\langle\Psi^{-}|.
\end{eqnarray}
For this case, Alice or Bob should make a local phase flip gate on her/his photon to obtain target state $|\Phi^{+}\rangle$.

\section{HHSZ+ and Simon-Pan protocols}\label{secHHSZ}
Considering HHSZ+ protocol \cite{husheng} in which a hyperentanglement in both polarization and spatial DoFs is used, the state of a system is $\rho_{in}=\rho_{p}\otimes\rho_{s}$. Here, $\rho_p$ and $\rho_s$ are the density operators of polarization and spatial DoFs, respectively. The Bell states of  spatial DoF are given by
\begin{eqnarray}\label{Bellspatial}\nonumber
|\Phi^{\pm}_s\rangle&=&\frac{1}{\sqrt{2}}(|0_{3}0_{4}\rangle\pm|1_{3}1_{4}\rangle),\\
|\Psi^{\pm}_s\rangle&=&\frac{1}{\sqrt{2}}(|0_{3}1_{4}\rangle\pm|1_{3}0_{4}\rangle).
\end{eqnarray}
As shown in Fig. \ref{HHSZ} (b), we use subscripts 1 (3) and 2 (4) to label the polarization (spatial) DoF of Alice and Bob, respectively.
Hyperentanglement is distributed to Alice and Bob, and will be a mixed one via a noisy channel given by
\begin{eqnarray}\label{mixedp}
\rho_{in}^{p}&=&f_{in}^{p}|\Phi^{+}_{p}\rangle\langle\Phi^{+}_{p}|+(1-f_{in}^{p})|\Psi^{+}_{p}\rangle\langle\Psi^{+}_{p}|,
\end{eqnarray}
and
\begin{eqnarray}\label{mixeds}
\rho_{in}^{s}&=&f_{in}^{s}|\Phi^{+}_{s}\rangle\langle\Phi^{+}_{s}|+(1-f_{in}^{s})|\Psi^{+}_{s}\rangle\langle\Psi^{+}_{s}|.
\end{eqnarray}
The coefficients $f_{in}^{p}$ and $f_{in}^{s}$ are the initial fidelities of polarization and spatial DoFs, respectively.
The fidelities satisfy conditions $f_{in}^{p}>\frac{1}{2}$ and $f_{in}^{s}>\frac{1}{2}$.
Therefore, the state of the whole system is
\begin{eqnarray}\label{}\nonumber
\rho_{1-4}&=&\rho_{12}\otimes\rho_{34}\\\nonumber
&=&\frac{1}{4}[f_{1}(|0_{1}0_{2}\rangle+|1_{1}1_{2}\rangle)(|0_{3}0_{4}\rangle+|1_{3}1_{4}\rangle)\langle...|\\\nonumber
&&+f_{2}(|0_{1}0_{2}\rangle+|1_{1}1_{2}\rangle)(|0_{3}1_{4}\rangle+|1_{3}0_{4}\rangle)\langle...|\\\nonumber
&&+f_{3}(|0_{1}1_{2}\rangle+|1_{1}0_{2}\rangle)(|0_{3}0_{4}\rangle+|1_{3}1_{4}\rangle)\langle...|\\\nonumber
&&+f_{4}(|0_{1}1_{2}\rangle+|1_{1}0_{2}\rangle)(|0_{3}1_{4}\rangle+|1_{3}0_{4}\rangle)\langle...|,
\end{eqnarray}
where the fidelities $f_{1}$, $f_{2}$, $f_{3}$ and $f_{4}$ are $f_{1}=f_{in}^{p}f_{in}^{s}$, $f_{2}=f_{in}^{p}(1-f_{in}^{s})$, $f_{3}=f_{in}^{s}(1-f_{in}^{p})$ and $f_{4}=(1-f_{in}^{p})(1-f_{in}^{s})$.
To purify above state, the bilateral CNOT gates realized by two PBSs are applied as
\begin{eqnarray}\label{CNOThyper}\nonumber
U^{PBS}_{CNOT}&=&\bigotimes^{2}_{i=1}(|0_{i}0_{i+2}\rangle\langle 0_{i}0_{i+2}|+|0_{i}1_{i+2}\rangle\langle 0_{i}1_{i+2}|\\
&&+|1_{i}1_{i+2}\rangle\langle 1_{i}0_{i+2}|+|1_{i}0_{i+2}\rangle\langle 1_{i}1_{i+2}|).
\end{eqnarray}
After the CNOT operations, the system is transferred to
\begin{eqnarray}\label{finalhyper}\nonumber
\rho^{cnot}_{1-4}&=&U^{PBS}_{CNOT}\rho_{1-4}U^{PBS\dagger}_{CNOT}\\\nonumber
&=&\frac{1}{4}[f_{1}(|0_{1}0_{2}0_{3}0_{4}\rangle+|1_{1}1_{2}1_{3}1_{4}\rangle\\\nonumber
&&+|0_{1}0_{2}1_{3}1_{4}\rangle+|1_{1}1_{2}0_{3}0_{4}\rangle)\langle...|\\\nonumber
&&+f_{2}(|0_{1}0_{2}0_{3}1_{4}\rangle+|1_{1}1_{2}1_{3}0_{4}\rangle\\\nonumber
&&+|0_{1}0_{2}1_{3}0_{4}\rangle+|1_{1}1_{2}0_{3}1_{4}\rangle)\langle...|\\\nonumber
&&+f_{3}(|0_{1}1_{2}0_{3}1_{4}\rangle+|1_{1}0_{2}1_{3}0_{4}\rangle\\\nonumber
&&+|0_{1}1_{2}1_{3}0_{4}\rangle+|1_{1}0_{2}0_{3}1_{4}\rangle)\langle...|\\\nonumber
&&+f_{4}(|0_{1}1_{2}0_{3}0_{4}\rangle+|1_{1}0_{2}1_{3}1_{4}\rangle\\
&&+|0_{1}1_{2}1_{3}1_{4}\rangle+|1_{1}0_{2}0_{3}0_{4}\rangle)\langle...|].
\end{eqnarray}
Analyzing the above density operator, we find that there are four cases, $0_{3}0_{4}$, $0_{3}1_{4}$, $1_{3}0_{4}$, and $1_{3}1_{4}$, for obtaining the final entangled pair. The cases $0_{3}0_{4}$ (two upper) and $1_{3}1_{4}$ (two lower) induce the system with $\rho_{out}^{p}=f_{in}^{p}f_{in}^{s}|\Phi^{+}_p\rangle\langle\Phi^{+}_p|+(1-f_{in}^{p})(1-f_{in}^{s})|\Psi^{+}_p\rangle\langle\Psi^{+}_p|$. The fidelity of the residual entangled pair $|\Phi^{+}_p\rangle$ is $f_{out}^{p}=\frac{f_{in}^{p}f_{in}^{s}}{f_{in}^{p}f_{in}^{s}+(1-f_{in}^{p})(1-f_{in}^{s})}$. With conditions $f_{in}^{p}>\frac{1}{2}$ and $f_{in}^{s}>\frac{1}{2}$, it has higher fidelity, i.e. $f_{out}^{p}>f_{in}^{p}$ and $f_{out}^{p}>f_{in}^{s}$.

In some special cases, such as strong robustness of spatial or time DoFs in experiments \cite{shengsb,ursinprapplied}, the distributed state of this robust DoF could be a pure state, e.g. $f_{in}^{s}=1$ in Eq. (\ref{mixeds}). The entanglement purification of this case is Simon-Pan protocol \cite{simonpan}. In Eq. (\ref{finalhyper}), if the spatial fidelity is $f_{in}^{s}=1$, the density operator after the CNOTs is
\begin{eqnarray}\label{finalhyperde}\nonumber
\rho^{cnot}_{1-4}&=&U^{PBS}_{CNOT}\rho_{1-4}U^{PBS\dagger}_{CNOT}\\\nonumber
&=&\frac{1}{4}[f^{p}_{in}(|0_{1}0_{2}0_{3}0_{4}\rangle+|1_{1}1_{2}1_{3}1_{4}\rangle\\\nonumber
&&+|0_{1}0_{2}1_{3}1_{4}\rangle+|1_{1}1_{2}0_{3}0_{4}\rangle)\langle...|\\\nonumber
&&+(1-f^{p}_{in})(|0_{1}1_{2}0_{3}1_{4}\rangle+|1_{1}0_{2}1_{3}0_{4}\rangle\\
&&+|0_{1}1_{2}1_{3}0_{4}\rangle+|1_{1}0_{2}0_{3}1_{4}\rangle)\langle...|].
\end{eqnarray}
One can see that the cases $0_{3}0_{4}$ and $1_{3}1_{4}$ with probability $f^{p}_{in}$ will produce $|\Phi^{+}_p\rangle$, and the $0_{3}1_{4}$ and $1_{3}0_{4}$ with probability $1-f^{p}_{in}$ obtain $|\Psi^{+}_p\rangle$ which can be transformed to $|\Phi^{+}_p\rangle$ by a local bit-flip operation. The fidelity of purified state is 1.

\section{Li and Sheng-Deng protocols}\label{secLi}
The entanglement purification using hyperentanglement for cases considering the both bit-flip and phase errors of polarization are Li \cite{lionestep} and Sheng-Deng \cite{shengonestep} protocols. When considering phase errors, the density operator will be added with the two terms of phase error and becomes
\begin{eqnarray}\label{EPLi}\nonumber
\rho_{in}^{p}&=&f_{in}^{p1}|\Phi^{+}_{p}\rangle\langle\Phi^{+}_{p}|+f_{in}^{p2}|\Psi^{+}_{p}\rangle\langle\Psi^{+}_{p}|\\
&&+f_{in}^{p3}|\Phi^{-}_{p}\rangle\langle\Phi^{-}_{p}|+f_{in}^{p4}|\Psi^{-}_{p}\rangle\langle\Psi^{-}_{p}|,
\end{eqnarray}
where the fidelities satisfy $f_{in}^{p4}=1-f_{in}^{p1}-f_{in}^{p2}-f_{in}^{p3}$. And the entangled spatial DoF is the ideal case $\rho_{s}=|\Phi^{+}_{s}\rangle\langle\Phi^{+}_{s}|$.
As shown in Figs. \ref{EPPLi} (a) and (b), the four half-wave plates used in the upper and lower paths act as CNOT gates whose control and target qubit are spatial and polarization DoFs, respectively. The two kinds of CNOT operations are
\begin{eqnarray}\label{CNOTEPLiu}\nonumber
U^{u}_{CNOT}&=&\bigotimes^{2}_{i=1}(|1_{i}0_{i+2}\rangle\langle 0_{i}0_{i+2}|+|0_{i}1_{i+2}\rangle\langle 0_{i}1_{i+2}|\\
&&+|0_{i}0_{i+2}\rangle\langle 1_{i}0_{i+2}|+|1_{i}1_{i+2}\rangle\langle 1_{i}1_{i+2}|),
\end{eqnarray}
and
\begin{eqnarray}\label{CNOTEPLid}\nonumber
U^{l}_{CNOT}&=&\bigotimes^{2}_{i=1}(|0_{i}0_{i+2}\rangle\langle 0_{i}0_{i+2}|+|1_{i}1_{i+2}\rangle\langle 0_{i}1_{i+2}|\\
&&+|1_{i}0_{i+2}\rangle\langle 1_{i}0_{i+2}|+|0_{i}1_{i+2}\rangle\langle 1_{i}1_{i+2}|).
\end{eqnarray}
Here, $U^{u}_{CNOT}$ and $U^{l}_{CNOT}$ are corresponding to Pauli-X gates (half-wave plates) in upper and lower paths, respectively.
So, with a series of operations in sequence of bilateral upper-CNOT, PBS-CNOT and lower-CNOT gates shown in Figs. \ref{EPPLi} (a) and (b), the entangled pair is governed by $U_{lpu}$ as follows:
\begin{eqnarray}\label{CNOTdpu}\nonumber
U_{lpu}&=&U^{l}_{CNOT}U^{PBS}_{CNOT}U^{u}_{CNOT}\\\nonumber
&=&\bigotimes^{2}_{i=1}(|0_{i}0_{i+2}\rangle\langle 1_{i}0_{i+2}|+|1_{i}1_{i+2}\rangle\langle 0_{i}1_{i+2}|\\
&&+|1_{i}0_{i+2}\rangle\langle 1_{i}1_{i+2}|+|0_{i}1_{i+2}\rangle\langle 0_{i}0_{i+2}|).
\end{eqnarray}
The final state of system is calculated as
\begin{eqnarray}\label{CNOTEPLifinal}\nonumber
\rho&=&U_{lpu}(\rho_{p}\otimes\rho_{s})U^{\dagger}_{lpu}
=f_{in}^{p1}(|0_{1}0_{2}1_{3}1_{4}\rangle+|0_{1}0_{2}0_{3}0_{4}\rangle\\\nonumber
&&+|1_{1}1_{2}1_{3}1_{4}\rangle+|1_{1}1_{2}0_{3}0_{4}\rangle)\langle...|\\\nonumber
&&+f_{in}^{p2}(|0_{1}0_{2}1_{3}0_{4}\rangle+|0_{1}0_{2}0_{3}1_{4}\rangle\\\nonumber
&&+|1_{1}1_{2}1_{3}0_{4}\rangle+|1_{1}1_{2}0_{3}1_{4}\rangle)\langle...|\\\nonumber
&&+f_{in}^{p3}(|0_{1}0_{2}1_{3}1_{4}\rangle-|0_{1}0_{2}0_{3}0_{4}\rangle\\\nonumber
&&+|1_{1}1_{2}1_{3}1_{4}\rangle-|1_{1}1_{2}0_{3}0_{4}\rangle)\langle...|\\\nonumber
&&+f_{in}^{p4}(|0_{1}0_{2}1_{3}0_{4}\rangle-|0_{1}0_{2}0_{3}1_{4}\rangle\\
&&+|1_{1}1_{2}1_{3}0_{4}\rangle-|1_{1}1_{2}0_{3}1_{4}\rangle)\langle...|.
\end{eqnarray}
Analyzing the above density operator, one can find that all the four cases of spatial DoF, i.e. $0_{3}0_{4}$, $0_{3}1_{4}$, $1_{3}0_{4}$, and $1_{3}1_{4}$ will obtain $|\Phi^{+}_p\rangle$ with fidelity 1.

The learning results show that the three gates arranged with the sequences of p-s-CNOT (p is control bit and s is target) $\otimes$ s-p-CNOT (s is control bit and p is target) $\otimes$ p-s-CNOT and s-p-CNOT $\otimes$ p-s-CNOT $\otimes$ s-p-CNOT can realize the same goal with the Li protocol. Some kinds of gate combinations are given as follows.
The l-p-l gate is
\begin{eqnarray}\label{CNOTdpd}\nonumber
U_{lpl}&=&U^{l}_{CNOT}U^{PBS}_{CNOT}U^{l}_{CNOT}\\\nonumber
&=&\bigotimes^{2}_{i=1}(|0_{i}0_{i+2}\rangle\langle 0_{i}0_{i+2}|+|1_{i}1_{i+2}\rangle\langle 1_{i}1_{i+2}|\\
&&+|1_{i}0_{i+2}\rangle\langle 0_{i}1_{i+2}|+|0_{i}1_{i+2}\rangle\langle 1_{i}0_{i+2}|).
\end{eqnarray}
The above l-p-l gate equals to p-l-p gate, i.e., $U_{plp}=U^{PBS}_{CNOT}U^{l}_{CNOT}U^{PBS}_{CNOT}=U_{lpl}$. The p-l-p gate is shown as an example in Fig. \ref{EPPLi} (c) and (d). The u-p-u gate is expressed by
\begin{eqnarray}\label{CNOTupu}\nonumber
U_{upu}&=&U^{u}_{CNOT}U^{PBS}_{CNOT}U^{u}_{CNOT}\\\nonumber
&=&\bigotimes^{2}_{i=1}(|1_{i}0_{i+2}\rangle\langle 1_{i}0_{i+2}|+|0_{i}1_{i+2}\rangle\langle 0_{i}1_{i+2}|\\
&&+|0_{i}0_{i+2}\rangle\langle 1_{i}1_{i+2}|+|1_{i}1_{i+2}\rangle\langle 0_{i}0_{i+2}|).
\end{eqnarray}
It is also equal to p-u-p gate as $U_{pup}=U^{PBS}_{CNOT}U^{u}_{CNOT}U^{PBS}_{CNOT}=U_{upu}$. Another gate u-p-l is written by
\begin{eqnarray}\label{CNOTupd}\nonumber
U_{upl}&=&U^{u}_{CNOT}U^{PBS}_{CNOT}U^{l}_{CNOT}\\\nonumber
&=&\bigotimes^{2}_{i=1}(|1_{i}0_{i+2}\rangle\langle 0_{i}0_{i+2}|+|0_{i}1_{i+2}\rangle\langle 1_{i}1_{i+2}|\\
&&+|0_{i}0_{i+2}\rangle\langle 0_{i}1_{i+2}|+|1_{i}1_{i+2}\rangle\langle 1_{i}0_{i+2}|).
\end{eqnarray}


\end{document}